# A REVIEW OF THE ENERGY EFFICIENT AND SECURE MULTICAST ROUTING PROTOCOLS FOR MOBILE AD HOC NETWORKS

Busola S. Olagbegi[1] and Natarajan Meghanathan[2]

[1,2]Jackson State University, 1400 Lynch St, Jackson, MS, USA
[1]shale_o@yahoo.com, [2]natarajan.meghanathan@jsums.edu

## ABSTRACT

*This paper presents a thorough survey of recent work addressing energy efficient multicast routing protocols and secure multicast routing protocols in Mobile Ad hoc Networks (MANETs). There are so many issues and solutions which witness the need of energy management and security in ad hoc wireless networks. The objective of a multicast routing protocol for MANETs is to support the propagation of data from a sender to all the receivers of a multicast group while trying to use the available bandwidth efficiently in the presence of frequent topology changes. Multicasting can improve the efficiency of the wireless link when sending multiple copies of messages by exploiting the inherent broadcast property of wireless transmission. Secure multicast routing plays a significant role in MANETs. However, offering energy efficient and secure multicast routing is a difficult and challenging task. In recent years, various multicast routing protocols have been proposed for MANETs. These protocols have distinguishing features and use different mechanisms.*

## 1. INTRODUCTION

An ad hoc network consists of a collection of autonomous mobile nodes formed by means of multi-hop wireless communication without the use of any existing network infrastructure. Ad hoc networks have become increasingly relevant in recent years due to their potential applications in battlefield, emergency disaster relief and etc. In an ad hoc network, each mobile node can serve as a router. A mobile ad-hoc network (MANET) is characterized by mobile nodes without any infrastructure. Mobile nodes self-organize to form a network over radio links. The goal of MANETs is to broaden mobility into the area of autonomous, mobile and wireless domains, where a set of nodes form the network routing infrastructure in an ad-hoc manner. This emerging trend has stirred the support of applications which range from highly dynamic Vehicular ad hoc networks (VANETs) to less dynamic applications such as moderately mobile peer-to-peer wireless networks.

In ad hoc networks, nodes communicate with each other by way of radio signals, which are broadcast in nature. Broadcast is a unique case of multicast, wherein all nodes in the network should get the broadcast message. Multicasting is a communication process in which the transmission of packets (message) is initiated by a single user and the message is received by one or more end users of the network. Multicasting in wired and wireless networks has been advantageous and used as a vital technology in many applications such as audio/ video conferencing, corporate communications, collaborative and groupware applications, distance earning, stock quotes, distribution of software, news and etc [1]. Under multicast communications, a single stream of data can be shared with multiple recipients and data is only duplicated when required [1].

In the wired settings, there are two popular multicast tree schemes: shortest-path tree and core-based tree. The procedure to construct shortest-path multicast trees ensures the shortest path from every source to every destination, but a source node has to construct a tree rooted at itself. Hence, there would exist too many shortest-path trees existing in the network. In core-based multicast trees, shortest path from the source node to the destination node cannot be guaranteed, but only one tree would be needed to connect the set of the source nodes to a set of the receiver nodes.

Security is a more sensitive issue in MANETs than any other networks due to lack of infrastructure and the broadcast nature of the network. While MANETs can be quickly set up as needed, they also need secure routing protocols to add the security feature to normal routing protocols. The need for more effective security measures arises as many passive and active security attacks can be launched from the outside by malicious hosts or from the inside by compromised nodes [2]. Key management is a fundamental part of secure routing protocols; existence of an effective key management framework is also paramount for secure routing protocols. Several security protocols have been proposed for MANETs, there is no approach fitting all networks, because the nodes can vary between any devices.

However, it would be a difficult and challenging task to offer energy efficient and reliable multicast routing in MANETs. It might not be possible to recharge / replace a mobile node that is powered by batteries during a mission. The inadequate battery lifetime imposes a limitation on the network performance. To take full advantage of the lifetime of nodes, traffic should be routed in a way that energy consumption is minimized. In recent years, various energy efficient multicast routing protocols have been proposed. These protocols have unique attributes and utilize different recovery mechanisms on energy consumption. This project will provide a comprehensive understanding of these multicast routing protocols and better organize existing ideas and work to make it easy to design multicast routing in MANETs. The goal of this paper is to help researchers to gain a better understanding of energy-efficient and secure routing protocols available and assist them in the selection of the right protocol for their work. The rest of the paper is organized as follows: Section 2 presents related work on comparisons and surveys of multicast routing protocols for MANETs. Section 3 describes the energy-efficient multicast routing protocols and Section 4 describes the security-based multicast routing protocols surveyed for MANETs. Section 5 concludes the paper.

## 2. RELATED WORK

MANETs are gaining tremendous focus from researchers and application developers because of the great potential of its network type. Some research has been made in the field of multicast routing protocols; the taxonomy, performance and capacity of multicast routing protocols over MANETs have been studied ([3], [4]). Omari et al summarized traffic models for multicast routing protocols in MANETs ([5], [6]). They also evaluated the performance of the existing multicast protocols in MANET using similar traffic models to justify their proposal. Multicast routing protocols were categorized into tree-based mesh-based, stateless, hybrid-based and flooding protocols.

Four distinctive multicast routing protocols were discussed in detail, with a focus on how to rise above the constraints present in the previously proposed multicast protocols. The four multicast routing protocols discussed in [6] are: On-Demand Multicast Routing Protocol (ODMRP), Multicast Ad Hoc On-Demand Distance Vector Routing Protocol (MAODV), Forwarding Group Multicast Protocol (FGMP) and Core-Assisted Mesh Protocol. Chen et al gave a general survey of multicast routing protocols in MANETs [1] and called attention to the constraints, including Quality of Service (QoS) and reliability, faced in the design of these protocols when they are applied in highly dynamic environments, characteristics of MANETs.

QoS is difficult to guarantee in MANETs, due to lack of resources from sharing wireless bandwidth among nodes and topology changes as the nodes move. Protocols such as Location guided Tree (LGT) [7], Ad hoc Multicast Routing protocol utilizing Increasing id-numberS (AMRIS) [8], [9] and Core-Assisted Mesh Protocol (CAMP) [10] may not be realistic to use, because QoS was not considered in the design as it requires finding a route from a source to a destination and satisfying the end-to-end QoS requirement which is usually given in terms of bandwidth or delay. Proper QoS cost metrics such as bandwidth, delay, packet loss rate should be used in the design of multicast routing protocols [6].

The MAODV protocol constructs multicast trees to reduce end-to-end latency while ODMRP constructs a multicast mesh to guarantee robustness. The Position Based Multicast (PBM) protocol neither constructs a tree nor a mesh; it uses the geographic position of the nodes to make forwarding decisions. The Progressively Adapted Sub-Tree in Dynamic Mesh (PAST-DM) protocol builds a virtual mesh spanning all the members of a multicast group. In order to transmit and deliver packets, it depends on the underlying unicast routing protocol, leading to longer delays and lower on packet delivery [11]. Dewan also introduced a new protocol; Lifetime – Refining Energy Efficient of Multicast Trees (LREMiT) [12] that aims to maximize the lifetime of the multicast tree through refinement operation. This operation continues in rounds coordinated by the source node. The Protocol for Unified Multicasting through Announcement (PUMA) [13] uses a set of core nodes for multicasting by creating and maintaining a shared mesh for each multicast group without depending upon a unicast routing protocol. PUMA delivers data at a higher efficiency, while also provides a tight bound for control overhead in a wide range of network scenarios.

The Sequence and Topology encoding for Multicast Protocol (STMP) [14] represents multicast routing a using fuzzy Petri-net model in wireless ad hoc networks wherein all nodes are equipped with GPS units [15]. It uses a structured representation of network topology and applies a fuzzy reasoning algorithm in order to construct multicast tree and improves the efficiency of the routing protocol. Its main objective is to reduce the size of the multicast tree.

In Adaptive Demand Driven Multicast Routing (ADMR) [16], senders and receivers cooperate to establish and maintain forwarding states in the network to allow multicast communication [17]. ADMR adaptively monitors the proper execution of forwarding states and maintains connectivity when one or more forwarding nodes or receivers become disconnected. The Lifetime-aware Multicast Tree (LMT) [18] routing algorithm maximizes multicast lifetime by finding routing solutions that minimizes the variance of the available energy levels in the network [8].

Prioritized Overlay Multicast (POM) [19] aims to improve the efficiency and robustness of the overlay multicast in MANETs by building multiple role-based prioritized trees [20]. Usually it takes the benefits of location information. Cordeiro et al. provide information about the current state-of-the art in multicast protocols for MANETs, and compares them with respect to several performance metrics. In [21] and [22], the authors classify the multicast protocols into four categories based on the creation of the routes by the members of the group: tree-based approaches, meshed-based approaches, stateless multicast and hybrid approaches.

## 3. SURVEY OF ENERGY EFFICIENT MULTICAST ROUTING PROTOCOLS

MANETs lack fixed infrastructure and nodes are typically powered by batteries with a limited energy supply wherein each node stops functioning when the battery drains. Energy efficiency is an important consideration in such an environment. Since nodes in MANETs rely on limited battery power for their energy, energy-saving techniques aimed at minimizing the total power consumption of all nodes in the multicast group (minimize the number of nodes used to establish multicast connectivity, minimize the control overhead and so on) and at maximizing the multicast life span should be considered. As a result of the energy constraints placed on the network's nodes, designing energy efficient multicast routing protocols is a crucial concern for MANETs, to maximize the lifetime of its nodes and thus of the network itself [23], [24].

Energy-efficient broadcast routing algorithms called Minimum Longest Edge (MLE) and Minimum Weight Incremental Arborescence (MWIA) are introduced in [5][20]. MLE is able to achieve a longer network lifetime by reducing the maximum transmission power of nodes. With MLE, the likelihood that a node is overused is reduced significantly. This scheme was expanded by considering a scenario where we introduce edge weights on the basis of the remaining energy of the sending nodes and receiving nodes. MWIA was derived from this idea, which is the best possible solution for broadcast routing with the minimum largest edge-weight.

Cheng et al. proposed the Minimum Incremental Power (MIP) algorithm and it is known as the most energy-efficient heuristic in terms of the total energy consumption among all the topologies [17]. MIP is developed based on the Broadcast Incremental Power (BIP) algorithm. The MIP algorithm is used as a comparison for the solution to the Energy-balanced topology control problem, which instead of minimizing the total energy, minimizes the maximum energy consumption at each node.

### 3.1 Energy-Efficient Location Aided Routing (EELAR)

Energy Efficient Location Aided Routing (EELAR) Protocol [3] was developed on the basis of the Location Aided Routing (LAR) [25]. EELAR makes significant reduction in the energy consumption of the mobile node batteries by limiting the area of discovering a new route to a smaller zone. Thus, control packet overhead is significantly reduced. In EELAR, a reference wireless base station is used and the network's circular area centered at the base station is divided into six equal sub-areas. During route discovery, instead of flooding control packets to the whole network area, they are flooded to only the sub-area of the destination mobile node. The base station stores locations of the mobile nodes in a position table. Simulations results using NS-2 [26][27] showed that EELAR protocol makes an improvement in control packet overhead and delivery ratio compared to AODV [28], LAR [29], and DSR [30][31] protocols.

### 3.2 Online Max-Min Routing Protocol (OMM)

Li et al proposed the Online Max-Min (OMM) power-aware routing protocol [21] for wireless ad-hoc networks dispersed over large geographical areas to support applications where the message sequence is not known. This protocol optimizes the lifetime of the network as well as the lifetime of individual nodes by maximizing the minimal residual power, which helps to prevent the occurrence of overloaded nodes. In most applications that involve MANETs, power management is a real issue and can be done at two complementary levels (1) during communication and (2) during idle time. The OMM protocol maximizes the lifetime of the network without knowing the data generation rate in advance. The metrics developed showed that OMM had a good empirical competitive ratio to the optimal online algorithm [21] that knows the message sequence and the max-min achieves over 80% of the optimal node lifetime (where the sender knows all the messages ahead of time) for most instances and over 90% of the optimal node lifetime for many problem instances [5].

### 3.3 Power-aware Localized Routing (PLR)

The Power-aware Localized Routing (PLR) protocol [32] is a localized, fully distributed energy-aware routing algorithm but it assumes that a source node has the location information of its neighbors and the destination. PLR is equivalent to knowing the link costs from the source node to its neighbors, all the way to the destination. Based on this information, the source cannot find the optimal path but selects the next hop through which the overall transmission power to the destination is minimized [5].

### 3.4 Power-aware Routing (PAR) Protocol

Power-aware routing (PAR) [33] maximizes the network lifetime and minimizes the power consumption by selecting less congested and more stable route, during the source to destination route establishment process, to transfer real-time and non real-time traffic, hence providing energy efficient routes. PAR focuses on 3 parameters: Accumulated energy of a path, Status of battery lifetime and Type of data to be transferred. At the time route selection, PAR focuses on its core metrics like traffic level on the path, battery status of the path, and type of request from user side. With these factors in consideration, PAR always selects less congested and more stable routes for data delivery and can provide different routes for different type of data transfer and ultimately increases the network lifetime. Simulation results shows that PAR outperforms similar protocols such as DSR and AODV, with respects to different energy-related performance metrics even in high mobility scenarios. Although, PAR can somewhat incur increased latency during data transfer, it discover routed that can last for a long time and encounter significant power saving.

### 3.5 Minimum Energy Routing (MER) Protocol

Minimum Energy Routing (MER) can be described as the routing of a data-packet on a route that consumes the minimum amount of energy to get the packet to the destination which requires the knowledge of the cost of a link in terms of the energy expanded to successfully transfer and receive data packet over the link, the energy to discover routes and the energy lost to maintain routes [22]. MER incurs higher routing overhead, but lower total energy and can bring down the energy consumed of the simulated network within range of the theoretical minimum the case of static and low mobility networks. However as the mobility increases, the minimum energy routing protocol's performance degrades although it still yields impressive reductions in energy as compared performance of minimum hop routing protocol [34].

### 3.6 Lifetime-aware Multicast Tree (LMT) Protocol

The Lifetime-aware multicast tree routing algorithm [18] maximizes the ad hoc network lifetime by finding routes that minimize the variance of the remaining energies of the nodes in the network. LMT maximizes the lifetime of a source based multicast tree, assuming that the energy required to transmit a packet is directly proportional to the forwarding distance. Hence, LMT is said to be biased towards the bottleneck node. Extensive simulation results were provided to evaluate the performance of LMT with respect to a number of different metrics (i.e., two definitions of the network lifetime, the root mean square value of remaining energy, the packet delivery ratio, and the energy consumption per transmitted packet) in comparison to a variety of existing multicast routing algorithms and Least-cost Path Tree (LPT) [35][36]. These results clearly demonstrate the effectiveness of LMT over a wide range of simulated scenarios.

### 3.7 Lifetime-aware Refining Energy Efficiency of Multicast Trees (L-REMIT)

Lifetime of a multicast tree in terms of energy is the duration of the existence of the multicast service until a node dies due its lack of energy. L-REMIT [12] is a distributed protocol and is part of a group of protocols called REMIT (Refining Energy efficiency of Multicast Trees). It uses a minimum-weight spanning tree (MST) as the initial tree and improves its lifetime by switching children of a bottleneck node to another node in the tree. A multicast tree is obtained from the "refined" MST (after all possible refinements have been done) by pruning the tree to reach only multicast group nodes. L-REMiT is a distributed algorithm in the sense that each node gets only a local view of the tree and each node can independently switch its parent as long as the multicast tree remains connected that utilizes an energy consumption model for wireless communication. L-REMiT takes into account the energy losses due to radio transmission as well as transceiver electronics. L-REMiT adapts a given multicast tree to a wide range of wireless networks irrespective of whether they use long-range radios or short-range radios [1], [12].

### 3.8 Localized Energy-aware Routing (LEAR) Protocol

Local Energy-Aware Routing (LEAR) [37] simultaneously optimizes trade-off between balanced energy consumption and minimum routing delay and also avoids the blocking and route cache problems. LEAR accomplishes balanced energy consumption based only on local information, thus removes the blocking property. Based on the simplicity of LEAR, it can be easily be integrated into existing ad hoc routing algorithms without affecting other layers of communication protocols. Simulation results show that energy usage is better distributed with the LEAR algorithm as much as 35% better compared to the DSR algorithm. LEAR is the first protocol to explore balanced energy consumption in a pragmatic environment where routing algorithms, mobility and radio propagation models are all considered [5], [12].

### 3.9 Conditional Max-Min Battery Capacity Routing (CMMBCR) Protocol

The Conditional Max-Min battery capacity routing (CMMBCR) [38] protocol utilizes the idea of a threshold to maximize the lifetime of each node and to fairly use the battery fairly. If all nodes in some possible routes between a source-destination pair have larger remaining battery energy than the threshold, the min-power route among those routes is chosen [5]. If all possible routes have nodes with lower battery capacity than the threshold, the max-min route is chosen. CMMBCR protocol selects the shortest path if all nodes in all possible routes have adequate battery capacity (i.e. the greater threshold). When the battery capacity for some nodes goes below a predefined threshold, routes going through these nodes will be avoided, and therefore the time until the first node failure, due to the exhaustion of battery capacity is extended. By adjusting the value of the threshold, we can maximize either the time when the first node powers down or the lifetime of most nodes in the network [39].

### 3.10 SPAN: An Energy Efficient Coordination Algorithm for Topology Maintenance

SPAN: An energy-efficient coordination algorithm for topology maintenance [40] is a distributed synchronization technique for multi hop ad hoc wireless networks that minimizes energy consumption without notably diminishing the connectivity of the network. SPAN coordinates the "stay-awake and sleep" cycle of the nodes and also performs multi-hop packet routing within the ad hoc network, while other nodes remain in power saving mode and periodically check if they should remain awaken and become a coordinator. SPAN adaptively elects coordinators by allowing each node to use a random back-off delay to decide whether to become a coordinator in the network and rotates them in time. The back-off delay for a node is a function of the number of other nodes in the neighbourhood and the amount of energy left in these nodes. This technique not only preserves network connectivity, it also preserves capacity, decreases latency and provides significant energy savings. The amount of energy saving provided by SPAN increases only slightly as density decreases. Current implementation of span uses the power saving features, since the nodes practically wake up and listen for traffic advertisements [5].

### 3.11 Power-aware Multiple Access (PAMAS) Protocol

PAMAS [41] is an extension to the AODV protocol; it uses a new routing cost model to discourage the use of nodes running low on battery power. PAMAS also saves energy by turning off radios when the nodes are not in use. Results show that the lifetime of the network is improved significantly. There is a trivial negative effect on packet delivery fraction and delay, except at high traffic scenarios, where both actually improve due to reduced congestion.

Routing load, however, is consistently high, more at low traffic scenarios. For the most part, PAMAS demonstrates significant benefits at high traffic and not-so-high mobility scenarios. Although, it was implemented on the AODV protocol, the technique used is very standard and can be used with any on-demand protocol. The energy-aware protocol works only in the routing layer and exploits only routing-specific information [13].

### 3.12 Geographic Adaptive Fidelity (GAF) Protocol

Geographical adaptive fidelity (GAF) protocol [34], [42] reduces energy consumption in ad hoc wireless networks; it is used for extending the lifetime of self-configuring systems by exploiting redundancy to conserve energy while maintaining application fidelity. By identifying nodes that are equivalent from a routing perspective and then turning off unnecessary nodes, maintaining a constant level of routing fidelity, this protocol is able to conserve energy. GAF also uses application-and system-level information; nodes that source or sink data remain on and intermediate nodes monitor and balance energy use. GAF is independent of the underlying ad hoc routing protocol; simulation studies of GAF show that it can consume 40% to 60% less energy than other ad hoc routing protocol. Also, network lifetime increases proportionally to node density [5].

### 3.13 Prototype Embedded Network (PEN) Protocol

The Prototype Embedded Network (PEN) protocol [23] exploits the low duty cycle of communication activities and powers down the radio device when it is idle. Nodes interact asynchronously without master nodes and thus, the costly master selection procedure as well as the master overloading problem can be avoided. But in order for nodes to communicate without a central coordinator, each node has to periodically wake up, make its presence by broadcasting beacons, and listens a moment for any communication request before powering down again. A transmitting source node waits until it hears a beacon signal from the intended receiver or server node. Then, it informs its intention of communication during the listening period of the server and starts the communication. Due to its asynchronous operation, the PEN protocol minimizes the amount of active time and thus saves substantial energy. However, the PEN protocol is effective only when the rate of interaction is fairly low, thus more suited for applications involving simple command traffic rather than large data traffic [5].

### 3.14 Protocol for Unified Multicast through Announcements (PUMA)

PUMA [13] is a protocol that uses simple multicast announcements to elect a core for the group and inform all routers of their distance and next-hops to the core, join, and leave the multicast group. PUMA provides the lowest and a very tight bound for the control overhead compared to ODMRP and MAODV. In other words, the control overhead of PUMA is almost constant node when mobility, number of senders, multicast group size or traffic load is changed. It also provides the highest packet delivery ratio for all scenarios [5]. The mesh constructed by PUMA provides redundancy to the region containing receivers, thus reducing unnecessary transmissions of multicast data packets. PUMA does not depend on the existence of any specific pre-assigned unicast protocol [1].

### 3.15 Predictive Energy-efficient Multicast Algorithm (PEMA)

The Predictive Energy-efficient Multicast Algorithm (PEMA) [43] exploits statistical properties of the network to solve scalability and overhead issues caused by large scale MANETs as opposed to relying on route details or network topology. The running time of PEMA depends on the multicast group size, not network size; this makes PEMA fast enough even for MANETs consisting of 1000 or more nodes. Simulation results show that PEMA not only results in significant energy savings compared to other existing algorithms, but also attains good packet delivery ratio in mobile environments. A distinct feature in PEMA is its speed; it is extremely fast because its running time is independent of its network size and the routing decision does not rely on the information about network topology or route details [33].

## 4. SURVEY OF SECURITY MULTICAST ROUTING PROTOCOLS

Secure communication is a major concern in wireless ad hoc networks due to the broadcast nature of this type of network, the existence of a wireless medium, and the lack of any centralized infrastructure. This makes MANETs vulnerable to eavesdropping, interference, spoofing, and etc. Multicast routing protocols should take this into account, especially because some of these protocols are applied in areas such as military (battlefield) operations, national crises, and emergency operations. The unique characteristics of MANETs, combined with security threats, demand solutions for securing ad hoc networks prior to their use in commercial and military applications. Some of the unique characteristics of MANETs that pose a strong challenge to the design of the secure multicast routing protocols include: open peer-to-peer network architecture, shared wireless medium, demanding resource constraints, and dynamic network topology. These challenges clearly make a case for designing and developing strong secure routing protocols that achieve extensive protection as well as provide desirable network performance [44].

Routing is a challenging aspect of moving packets around in a network. It is a significant problem because any node can perform the role of the router in MANET and security concepts were not included into the routing protocols when they were designed. It is important because the routing table forms the basis of the network operations and any corruption of routing table may lead to significant consequences. Routing attacks in mobile ad hoc network are more challenging since routing relies on the trustworthiness of all the nodes involved. Also it is difficult to differentiate between a compromised node from a selfish node, the latter of the nodes do not cooperate during packet forwarding mainly due to reduce availability of critical resources such as residual energy and available bandwidth [45].

Secure multicast routing protocols are complex to design, due to the fact that they operate with limited resources as well as due to the highly dynamic nature of the network. Existing routing protocols that are not secure are likely to spread routing information quickly as scenarios change, which causes more recurrent communication between the nodes in a conventional network. To avoid possible Denial of Service Attacks, resource-intensive and complicated security mechanisms to have to be deployed; but this can delay or prevent the network from exchanging information. Routing protocols for ad hoc networks generally can be divided into three main categories: Proactive protocols, Reactive protocols and Hybrid protocols.

In a proactive routing protocol, nodes from time to time exchange routing information with other nodes so that each node always knows a currently available route to all destinations. Maintenance of such an updated routing table requires each node to periodically broadcast control packets throughout the entire network. This means that the routes to destination nodes are calculated at regular time instants, before ascertaining the connection from source to destination. When a source node wants to send data to a destination node, it searches the routing table to find a route to the destination node. With the proactive routing approach, the control overhead is significantly large because control packets are sent periodically to maintain all routes even though some of the routes might not be used. An example of proactive routing protocol is Dynamic Sequenced Destination Vector (DSDV) routing protocol [46].

In contrast, in a reactive routing protocol, nodes exchange routing information only as needed; a node will only try to find out a route to some destination only when it has data packets to transmit to that destination (e.g., [**Error! Reference source not found.**]). These protocols usually have to go through two phases: route discovery (establish routes to destination) and route maintenance. The route to the destination could break, due to the dynamically changing topology, leading to frequent execution of route discovery and route maintenance procedures. This leads to an appreciable route acquisition latency that significantly contributes to the end-to-end delay per transmitted data packet. An example of a reactive routing protocol is Dynamic Source Routing Protocol (DSR) [47].

Hybrid routing protocols [48] is a combination of both proactive and reactive protocols. They use distance-vectors to precisely keep track of the routing metrics and to determine the best route to destination networks, and will broadcast routing table updates only when there is a change in the network topology. This allows for rapid convergence, and also requires reduced processing power and lower memory. An example of Hybrid routing protocols is the Zone Routing Protocol (ZRP) [49].

### 4.1 Security-aware Ad hoc Routing (SAR) Protocol

Security-aware Ad hoc Routing (SAR) [50] is a technique that incorporates security attributes as parameters into ad hoc route discovery. SAR uses of security as a metric to tradeoff and improve the applicability of the routes uncovered by ad hoc routing protocols. Ad hoc routing protocols enable nodes to mutually exchange information. Intermediate nodes receive data packets at a particular security level and process these packets or forward the packets relying on the security level of the intermediate node. If the required security level cannot be guaranteed, Route Request (RREQ) packets are dropped. Otherwise Route Reply (RREP) packets are sent back to the source from destination or intermediate nodes. This approach is resource demanding but it is a useful mechanism for prevention of attacks.

### 4.2 Security Efficient Ad hoc Distance Vector Routing Protocol (SEAD)

The Secure Efficient Ad Hoc Distance (SEAD) [24] routing protocol is a secure robust routing protocol that protects against multiple attackers creating incorrect routing state in any other node, even when active attackers are present in the network. The design of SEAD, in part, was based on the DSDV ad hoc network routing protocol, and especially, on the DSDV-SQ (for sequence number) version of the protocol. DSDV-SQ, is known to do better than other DSDV versions in terms of packet delivery ratio, although it does create more overhead in the network, due to an increase in the number and size of routing advertisements. The increase in the size of each advertisement is attributed to the addition of the hash value on each entry for authentication [44]. As a result, SEAD is efficient and can be used in energy and bandwidth-constrained nodes.

### 4.3 Security Grid Location Service Forwarding (SGLSF)

The SGLSF [51] mechanism was proposed after combining Secure Geographic Forwarding (SGF) [52] and Grid Location Service (GLS) [53]. The SGF mechanism uses the shared key and the Timed Efficient Stream Loss-tolerant Authentication (TIK) [54] protocol to provide source authentication, neighbor authentication, and message integrity by incorporating hashed message authentication code (MAC1). By combining these SGF and GLS, SGLSF, enhances the security to the original protocol to ensure that any receiver can authenticate the accuracy of location messages. SGLS has the ability to message tampering, dropping, falsified injection, and replay attacks. The authors in [51] also came up with another mechanism, called the Local Reputation System (LRS), and its aim is to discover compromised and selfish users and isolate messages by dropping attackers from the network. Both mechanisms combined; continue to maintain a larger message delivery ratio at the expense of a slightly higher average end-to-end delay and routing overhead compared to when they are not combined. After a performance analysis was done on both SGLS and LRS, and comparing them with the original GLS, results show that SGLS can operate efficiently by using effective cryptographic mechanisms.

### 4.4 Ariadne

Ariadne [55] is a secure on-demand routing protocol that prevents attackers or compromised nodes from interfering with uncompromised routes consisting of uncompromised nodes, and also prevents different of types of Denial-of-Service attacks. Araidne uses only highly efficient symmetric cryptographic primitives, which helps the secure protocol to effectively defend against node compromise. Araidne does not require any trusted hardware powerful processors. It can authenticate routing messages using one of three schemes: shared secrets between each pair of nodes, shared secrets between communicating nodes combined with broadcast authentication, or digital signatures [56], [57]. When Ariadne is paired with Timed Efficient Stream Loss-tolerant Authentication (TESLA) [58], [59], it provides an efficient broadcast authentication scheme that requires loose time synchronization. Use of pair-wise shared keys avoids the need for synchronization, but at the cost of higher key setup overhead; broadcast authentication such as TESLA also allows additional protocol optimizations.

### 4.5 EndairALoc

Secure routing protocols add security features to the normal routing protocols, leading to secure communication at the cost of increased energy consumption. Verification of the message authentication code and location information increases the computation consumption of the initiator and the latency of route discovery. The propagation of Route Request (RREQ) messages, each node takes a few actions. Intermediate nodes only need to generate message authentication codes during the propagation of Route Reply (RREP) messages. The EndairALoc protocol (an adaptation of Ariadne) provides the security of Araidne; but, also could resist the man-in-the-middle attack and even the wormhole attack. Furthermore, EndairALoc uses symmetric key mechanism instead of the public key mechanism, so energy consumption during route discovery decreases significantly.

### 4.6 Secure Routing Protocol (SRP)

Secure Routing Protocol (SRP) [60] is a secure route discovery protocol that mitigates the detrimental effects of malicious behavior and guarantees correct route discovery, so that altered, or replayed route replies are refused and will never get to the route requesting node. SRP only deduces a security association between the end-points of a path and so the nodes in between do not have to be trusted for route discovery. This is done by insisting that the Route Request (RREQ) along with a unique random query identifier reach the destination, where a route reply is built and a message authentication code is calculated over the path and returned to the source. The authors in [60] confirm the accuracy of the protocol analytically. SRP's responsiveness is safeguarded under different types of attacks that exploit the routing protocol itself. The sole requirement of this scheme is the existence of a security association between the node initiating the query and the sought destination; any two nodes that wish to communicate securely can simply establish a shared secret, to be used by their routing protocol modules.

### 4.7 Secure Link State Routing Protocol (SLSP)

Secure Link State Routing Protocol (SLSP) [61] is a proactive MANET protocol that secures the discovery and the distribution of link state information across mobile ad hoc domains. It provides accurate and authentic link state information, robustness against Byzantine behavior and failures of individual nodes. Byzantine Behavior can be described as any behavior in an ad hoc or wireless network where one or more devices work in collusion to disrupt the network. Examples are Denial of service behavior, dropping or altering packets, topology distortion, impersonation, wormholes, and a host of other security challenges. The scope of SLSP may range from a secure neighborhood discovery to a network-wide secure link state protocol. SLSP nodes disseminate their link state updates and maintain topological information for a zone of network nodes within R hops. More specifically, SLSP is capable of adjusting its scope between local and network-wide topology discovery and operating in networks of frequently changing topology and group membership.

### 4.8  Security-aware Adaptive DSR (SADSR) Protocol

The Security-aware Adaptive DSR (SADSR) [62] is a secure on-demand routing protocol for mobile ad hoc wireless networks in which every node maintains routes only with the nodes that it communicates with. SADSR authenticates the routing protocol messages using digital signatures based on asymmetric cryptography. The fundamental scheme behind SADSR is to manage multiple routes to each destination and store a local trust value for each node in the network. The trust value for a path is computed and assigned to each path based on trust values of the nodes which occur on the path. The paths with higher trust values are preferred for routing. When the performance of SADSR and DSR are compared, results show that in the presence of malicious nodes in the network SADSR outperforms DSR with respects to packet delivery ratio under an acceptable network load. In the presence of malicious nodes SADSR outperforms DSR with respect to throughput. In both cases, SADSR introduces a reasonable network load to establish high packet delivery ratio.

### 4.9  Authenticated Routing for Ad hoc Networks (ARAN)

Authenticated Routing for Ad hoc Networks (ARAN) [63] is a simple protocol that does not require significant additional work from nodes within the group by providing secure routing for the managed-open and open environments using public-key cryptographic mechanisms to defeat all identified attacks. Results show that ARAN is as effective as AODV in discovering and maintaining routes. The overhead with ARAN is large size routing packets, which result in a higher overall routing load, and higher latency in route discovery because of the cryptographic computation that must occur. ARAN can effectively and efficiently discover secure route in environments where nodes are authorized to participate but not trusted to cooperate, as well as environments where participants do not need to be authorized to participate. This scheme introduces authentication, message integrity, and non-repudiation to routing in an ad hoc environment as part of a minimal security policy. Results show that although there is a greater performance cost to ARAN as compared to DSR or AODV, the minimal increase in cost is outweighed by the increased security.

### 4.10  Secure Ad hoc On-Demand Distance Vector (SAODV) Routing Protocol

The Secure Ad hoc On-Demand Distance Vector (SAODV) [55] routing protocol is an efficient secure routing protocol for mobile ad hoc networks that guarantees the discovery of correct connectivity information over an unknown network, in the presence of malicious nodes. SAODV is capable of operating without the existence of an on-line certification authority or the complete knowledge of keys of all network nodes. Its sole requirement is that any two nodes that wish to communicate securely can simply establish a shared secret, to be used by their routing protocol modules. Moreover, the correctness of the protocol is retained irrespective of any permanent binding of nodes to IP addresses, a feature of increased importance for open, dynamic, and cooperative MANET environments.

## 5. CONCLUSIONS

A mobile ad hoc network (MANET) consists of autonomous mobile nodes, each of which communicates directly with the nodes within its wireless range or indirectly with other nodes in a network. In order to facilitate secure and reliable communication within a MANET, an efficient routing protocol is required to discover routes between mobile nodes. The field of MNAETs is rapidly growing due to the many advantages and different application areas. Energy efficiency and security are some challenges faced in MANETs, especially in designing a routing protocol. In this paper, we surveyed a number of energy efficient multicast routing protocols and secure multicast routing protocols. In many cases, it is difficult to compare these protocols with each other directly since each protocol has a different goal with different assumptions and employs mechanisms to achieve the goal. According to the study, these protocols have different strengths and drawbacks. A multicast protocol can hardly satisfy all requirements. In other words, one routing protocol cannot be a solution for all energy efficient and security issues that are faced in MANETs, but rather each protocol is designed to provide the maximum possible requirements, according to certain required scenarios.

In future years, as mobile computing keeps growing, MANETs will continue to flourish, and even if a multicast protocol meeting all the requirements it is designed for, it will be very complicated and will need a tremendous amount of routing information to be maintained. Moreover, it will not be suitable for environments with limited resources. Satisfying most of the requirements would provide support for secure communication, minimize storage and resource consumption, ensure optimal paths and minimize network load.

International journal on applications of graph theory in wireless ad hoc networks and sensor networks
(GRAPH-HOC) Vol.2, No.2, June 2010